# Coupling of Klein-Andreev Resonant States in Bi$_2$Sr$_2$CaCu$_2$O$_{8+x}$ -graphene- Bi$_2$Sr$_2$CaCu$_2$O$_{8+x}$ Devices

*Sharadh Jois[1,†], Jose L. Lado[3], Genda Gu[4], Qiang Li[4,5], *Ji Ung Lee[1,2]

[1] College of Nanoscale Science and Engineering, SUNY Polytechnic Institute, Albany, New York 12203, USA

[2] College of Nanotechnology, Science, and Engineering, University at Albany, SUNY, New York 12203, USA

[3] Department of Applied Physics, Aalto University, 00076 Aalto, Espoo, Finland

[4] Condensed Matter Physics and Materials Science Department, Brookhaven National Laboratory, Upton, New York, 11973, USA

[5] Department of Physics and Astronomy, Stony Brook University, Stony Brook, New York 11794, USA

## Abstract

Quantum devices require coherent coupling over macroscopic distances. Recently, resonances due to Klein tunneling and Andreev reflection states (KARS) have been observed in a naturally occurring *p-n* junction at the interface between Bi$_2$Sr$_2$CaCu$_2$O$_{8+x}$ (BSCCO), a high-*T$_c$* superconductor (HTS), and graphene. The resonances appear as conductance oscillations with gating. Here, we show coupling between the KARS in BSCCO-graphene-BSCCO devices of varying separation (L). The coupling is evidenced by a power-law decay of resonance period as L increases from tens of nanometers to single microns. These results demonstrate the long-distance coupling of KARS cavities in graphene-HTS junctions. The length dependence seen in experiments is supported by single-particle spectral functions which show KARS are coupled by transport modes in graphene. The strong coupling between KARS in BSCCO-graphene-BSCCO junctions showcases the novelty of HTS-graphene junctions for quantum circuits and unconventional Josephson junctions.

Main Text

Novel electronic resonant states can be seen in graphene (Gr) cavities, owing to its two-dimensional (2D) structure and Dirac dispersion [1]. Electronic cavities can occur naturally in graphene due to charge transfer [2,3] or be defined using electrostatic gating [4–6]. They can also occur between superconducting contacts in graphene Josephson junctions [7–10]. However, empirical evidence of hybridization between two resonant cavities in graphene is limited. Here, we demonstrate coupling between two resonant cavities that form in graphene and high-temperature superconductor (HTS) junctions based on weak power-law decay of resonance period as a function of junction length. This work will aid in creating unconventional Josephson junctions [11–14] using graphene HTS such as $Bi_2Sr_2CaCu_2O_{8+x}$ (BSCCO) and $YBa_2Cu_3O_7$ [15,16].

Recently, we reported on the formation of resonant cavities in single BSCCO-graphene junctions [17] due to two scattering mechanisms - Klein tunneling (KT) [18–20] and Andreev reflection (AR) [21,22], hereafter referred to as Klein-Andreev resonant states (KARS). KARS are confined to a *p*-doped region of graphene ($Gr'$) of length $d \approx 60$ nm that forms naturally at the boundary of a BSCCO-graphene junction due to hole doping from BSCCO ($n \approx 10^{15} cm^{-2}$) [22]. The $Gr'$ region has two interfaces. The first interface with the bare graphene region ($Gr$) contributes to Klein tunneling (KT), a relativistic tunneling effect that can occur in Dirac systems. This is supported by the reduced overall conductance we observed as a gate tunes this interface from an ambipolar *p-n* doping configuration to a unipolar *p-p* doping configuration. The second interface between the superconducting graphene region ($Gr_\Delta$) and $Gr'$ region contributes to Andreev reflection (AR). The $Gr_\Delta$ region forms due to the proximity effect from BSCCO. The suppression of resonances as the voltage across the junction increases above the size of the induced superconducting gap ($\Delta_i$) in graphene supports the superconducting origin of KARS. The single BSCCO-graphene junction with the $Gr_\Delta - Gr' - Gr$ graphene regions displays electronic resonances after two round-trips ($4 \times d$), similar to Fabry-Perot interference. This discovery prompts the investigation into BSCCO-graphene-BSCCO devices while controlling the separation between the two nearly identical $Gr'$ resonant cavities. Our transport studies discussed below show the length dependence of coupling between KARS cavities found in BSCCO-graphene-BSCCO devices.

Here, we fabricated two nearly identical KARS cavities in BSCCO-graphene-BSCCO devices with varied separation by breaking a single BSCCO and placing it on top of a continuous strip of graphene. We observe strong hybridization between the two $Gr'$ regions hosting KARS via the transport modes in the normal $Gr$ region. The coupling between the two junctions leads to an increased resonance period in short junctions. The resonance period follows a power-law decay with increasing separation length and approaches the single junction limit over a microscopic length separation. This observation confirms the hybridization between the two KARS cavities. In contrast, coupling between electronic cavities in gapped systems such as semiconductors or bilayer graphene will display an exponential decay. To support the experimental results and predictions, we calculated the spectral functions of the Gr' regions as we varied the length of the middle Gr region. Both experiments and calculations show a universal $L^c (c < 1)$ scaling in BSCCO-graphene-BSCCO junctions.

The schematic representation of the device and corresponding band diagram are shown in **Figure 1 (a)** having the form $Gr_{\Delta 1} - Gr'_1 - Gr - Gr'_2 - Gr_{\Delta 2}$. $Gr_{\Delta 1}$ and $Gr_{\Delta 2}$ are the graphene regions under BSCCO and form nodal superconducting regions due to the proximity effect. The two $p$-doped resonant regions $Gr'_1$ and $Gr'_2$ of lengths $d_1$ and $d_2$ form naturally at the boundary of each BSCCO$_{1,2}$/graphene junction due to charge transfer from BSCCO. The interfaces that form at the boundaries $Gr_{\Delta 1} - Gr'_1$ and $Gr_{\Delta 2} - Gr'_2$ contribute to AR. The other interface of each $p$-doped region, $Gr'_1 - Gr$ and $Gr'_2 - Gr$ interfaces, is responsible for KT. Thus two scattering processes, KT and AR, contribute to the resonances at each junction [4,23]. A global gate is used to sweep the Fermi level of the normal region, allowing us to detect the resonances in transport measurements. Negligible change to carrier density is expected in the superconducting regions of graphene due to screening effects from the very high charge transfer from BSCCO.

We fabricated several BSCCO-graphene-BSCCO devices with varying lengths ($L$) ranging from ~20 nm to ~3 µm. The experimental length dependence study of coupling between KARS is enabled by our fabrication process that keeps disorder to a minimum. Here, we modify the dry transfer technique used to pick-and-place layered materials by introducing lateral movements of the stamp to mechanically propagate a continuous crack in BSCCO while placing it on graphene [24]. This technique allows us to create BSCCO-graphene-BSCCO devices with nearly identical KARS cavities while allowing for the device length ($L$) to be varied. Each BSCCO-graphene junction in a given device is considered identical because they are made simultaneously from the same BSCCO flake. The sample images obtained from scanning electron microscopy (SEM) for the longest and shortest devices are shown in **Figures 1 (b)** and **(c)**, respectively. The images were collected after completing the transport measurements and are used to determine the average device lengths to aid in the analysis described below. Additionally, we developed a special etching technique so that contact resistances to BSCCO do not limit the transport measurements. The basic properties of each device can be found in *Table-S.1 of supplementary information*.

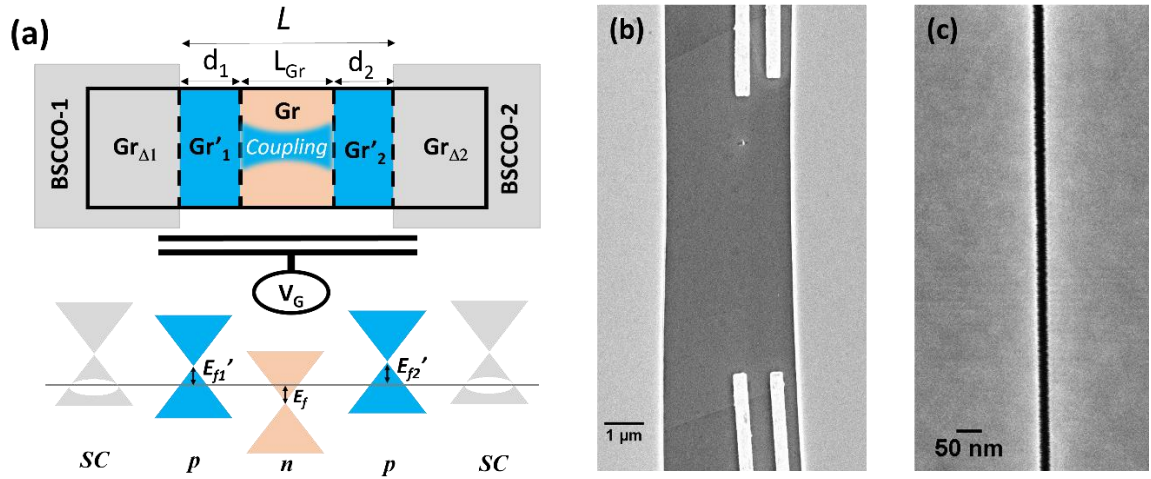

**Figure 1**. **(a)** Top panel: Schematic of S-Gr-S devices using BSCCO (S). Bottom panel: The band diagram of the different regions of graphene corresponding to the top panel. SC are regions of graphene with induced nodal superconductivity from BSCCO. The Gr' regions are the hole-doped resonant wells. The middle region of graphene ($Gr$) connects the two $Gr'_1$ and $Gr'_2$ regions and provides the medium for coupling the two KARS regions. Electron micrographs of long **(b)** and short **(c)** BSCCO-Gr-BSCCO devices.

We measure the four-terminal conductance across the BSCCO-Gr-BSCCO junctions at the liquid helium temperature as we vary the gate bias ($V_{bg}$) in steps of 0.1 V while holding a small, fixed bias of 1 mV across the junction. **Figure 2 (a)** shows the conductance of devices with different lengths. We stack the data vertically in different plots for clarity. Except for the $L = 0.047$ μm device measured at 20 K, all other devices were measured at temperatures ranging from 4.2 to 7 K in a probe station cooled with liquid helium. Based on our temperature studies in other devices, we anticipate similar results in the $L = 0.047$ μm device had we measured at lower temperatures. See for example *supplemental section* **Figure S.1** where we compare the data for the $L = 0.021$ μm device at 5 K and 30 K and show comparable resonance periods. Since the conductance at low temperature and bias in our devices is mostly oscillatory, the position of the Dirac point ($V_{DP}$) is determined from transfer curves taken at liquid nitrogen temperatures or for $V_{ds} > 50$ mV bias, well above twice the induced gap $Gr_{\Delta 1,2}$, which we determined from our single junction studies. The estimated range for uncertainty in $V_{DP}$ is given in ***Table-S1*** *of supplementary information.* In some devices, we inadvertently used substrates that used low carrier doping, resulting in carrier freeze-out for $V_{bg} < 0$. This issue was corrected in our earlier work [17] and for devices numbered S6 and above. Due to the carrier freeze-out, we are not able to report on the entire gate bias range. Therefore, we only report on the relevant regions, offset by $V_{DP}$, where carrier freeze-out was not an issue. The affected devices are numbered S5 and below in the *supplementary information*. The effect of carrier freezeout is shown in *supplemental* **Figure S.2** for the $L = 0.021$ μm device. The raw data is available at an online repository [25].

The same data in **Figure 2 (a)** is replotted with respect to Fermi-energy (E$_f$) in **Figure 2 (b)** using the relation $E_f = sign(V_{bg} - V_{DP})\hbar v_f \sqrt{\frac{\pi C_{bg}|V_{bg} - V_{DP}|}{e}}$, where, $\hbar$ is the reduced Planck constant, $v_f \sim 1 \times 10^6$ m/s is the Fermi velocity in graphene and $C_{bg} = 38.37$ nF/cm$^2$ is the gate capacitance for 90 nm SiO$_2$. For the $L = 0.268$ μm device and long devices ($L > 1$ μm), the oscillatory conductance is imposed on the background conductance modulation of the normal $Gr$ graphene region. The raw data from these four devices can found in the *supplemental **Figure S.3***. Therefore, for these devices, we show the oscillatory conductance (G$_{osc}$) after subtracting the background. The background of the measured signal is calculated using FFT or LOESS smoothing protocols in OriginLab analysis software. The G$_{osc}$ for these devices are also shown in *supplemental **Fig. S3***. In **Figure 2 (a)** and **(b),** we show the raw data for the rest of the devices without background subtraction. The resonances appear stronger as $L$ decreases below 1 μm, indicating stronger hybridization between the two KARS regions. In short devices, the uncertainty in determining the Dirac point ($V_{DP}$) is attributed to charge transfer doping effects from BSCCO. We account for this uncertainty as it contributes to errors in determining the primary resonance period shown later in **Figure 4**.

The present results are consistent with our single junction studies and further support the origin of AR as one of the scattering processes. Here, due to the formation of two junctions, we needed to apply twice the DC bias of the single junction device before observing a drop in the magnitude of the resonances. This is the voltage needed to diminish the effects of AR. For the present devices, this cut-off voltage is $V_{ds} = 20 - 30$ mV, about twice that of the single junction devices (see *supplemental section*, **Figures S.4 and S.5**) of $V_{ds} = 10 - 15$ mV, corresponding to the proximity-induced superconducting gap ($\Delta_i \sim 10 - 15 meV$) in graphene.

To analyze the data, we show in **Figure 2 (c)** the fast Fourier transform (FFT) amplitudes of the conductance for different devices as a function of the period (1/frequency). The shortest junction device ($L = 21$ nm), the top curve in red curve, shows oscillatory conductance with the largest period of 42 meV. For devices with $0.05 < L < 1$ μm, the primary period is between 25 to 35 meV. The dominant period in the long junctions ($L > 1$ μm) is visibly reduced to ~17 meV, approaching the single junction limit that we studied previously. The extrapolation to the single junction device limit clearly suggests that the two junctions are nearly identical and further supports the increasing hybridization effects between the KARS cavities as the length is reduced. As further support, see *supplemental* **Figure S. 6**, where we show that the period of the entire BSCCO-graphene-BSCCO device is comparable to the period of the individual BSCCO-graphene junctions measured on the same device. The trend of decreasing primary FFT period as a function of junction length is captured in **Figure 4**, showing a weak length dependence. Our theoretical calculations in **Figure 3** also show a similarly weak length dependence, confirming the hybridization between KARS in regions $Gr'_1$ and $Gr'_2$ through transport modes.

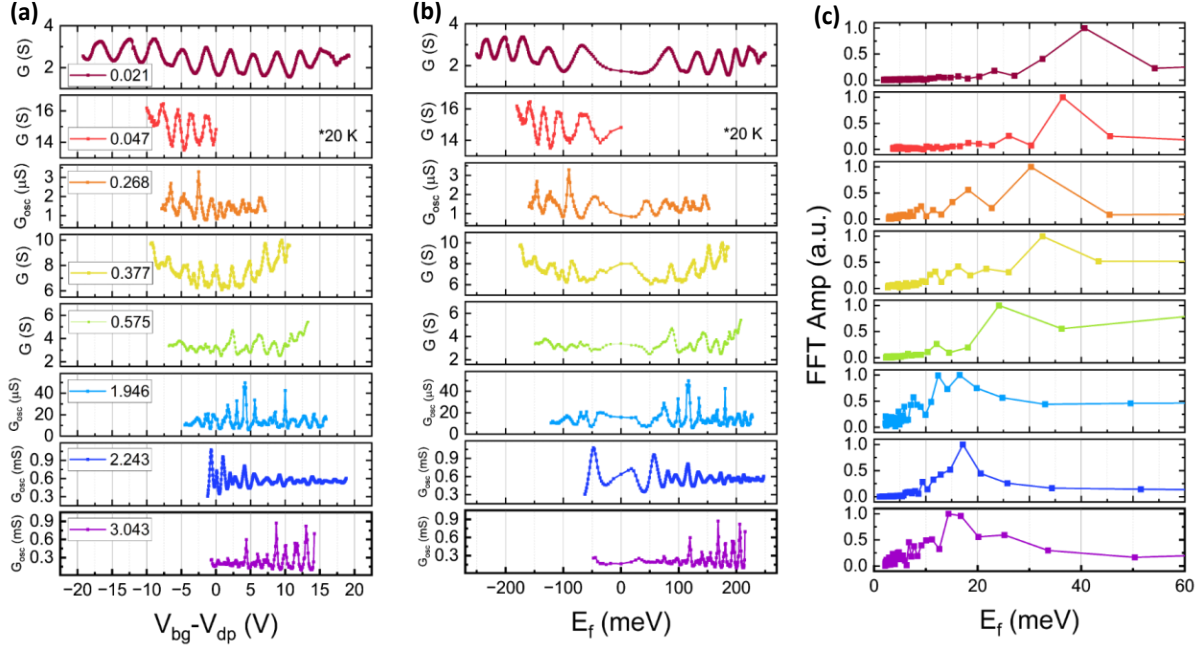

**Figure 2**. Normalized conductance in response to sweeping the back gate voltage ($V_{bg}$) of various devices is offset for comparison as a function of (a) gate voltage and (b) Fermi energy. The plot legends in (a) corresponds to the device length, L. The data was collected at liquid helium temperatures of 4.2 K for all devices, except for the device with L=0.047 μm, collected at 20 K, which should not affect the conclusion of the results. For junctions with $L > 1$ μm and $L = 0.268$ μm, the background conductance is subtracted to emphasize the oscillatory conductance ($G_{osc}$). The corresponding normalized FFT amplitudes of the resonances as function of 1/frequency, i.e., period in Fermi energy is shown in (c). Notice that the dominant period decreases as junction length increases. The respective junction lengths of the devices are given in the plot legends.

Additionally, we observe secondary FFT peaks, also seen in our calculations, which further supports the hybridization model of KARS over the electronic modes in the $Gr$ region. They generally follow the length dependence of the primary peaks. For devices with $L$ between $0.5 - 2$ μm, they appear around 15 meV, near the primary period of the single junction, showing the robustness of the hybridization over long distances. The secondary peaks progressively increase, appearing between 25 to 35 meV for shorter devices. Crucially, we observe higher order peaks for the longest device, which we attribute to disorder as we discuss below.

Not all higher order peaks are due to resonant modes, particularly those seen in longer devices. In particular, we observe noisy fluctuations in the longest device with periods below 10 meV. These can be attributed to universal conductance fluctuations owing to carrier scattering from charge puddles and disorder. The near absence of higher order peaks in the shorter devices suggests that disorder is minimal.

To support the experimental results, we calculated the electronic spectral functions of the device. The details of the modeling framework are described in [17]. We show the spectral functions of a $Gr'$ region hosting KARS with respect to the transverse momenta as the length of the middle $Gr$ region is varied in the $Gr_\Delta - Gr' - Gr - Gr' - Gr_\Delta$ junction, increasing the separation by $\left(\frac{L_{Gr}}{d}\right) = \frac{1}{2}, 1$, and 2 respectively in **Figures 3 (a), (b)** and **(c)**. $L_{Gr}$ and $d$ are lengths of the $Gr$ and $Gr'$ regions, respectively. The two $Gr'$ regions are treated identically. For comparison, the spectral function for an isolated Gr' region for a single $Gr_\Delta - Gr' - Gr$ junction is given in **Figure 3 (d)**, showing a qualitatively different resonant condition to the two-junction case. The insets in **Figure 3 (a-d)** show the schematic of the device structure as the length of the $Gr$ is increased. For the shortest junction with $L_{Gr} = \frac{1}{2}d$, shown in **Figure 3 (a)**, and focusing on K=0, one observes a noticeable separation in the resonant states, indicative of strong hybridization between the two $Gr'$ KARS cavities. This should be compared to the results with longer $L_{Gr}$ (**Figure 3 (b)** and **(c)**), where we observe noticeable decoupling between the KARS regions, approaching that of the single junction case of **Figure 3 (d)**. This trend is also reproduced in our experiment. The scaling from the calculations is captured in **Figures 3 (e)** showing the spectral function at zero-momentum as a function of $\frac{L_{Gr}}{d}$. Further analysis of change in resonance periodicity due to coupling is presented below.

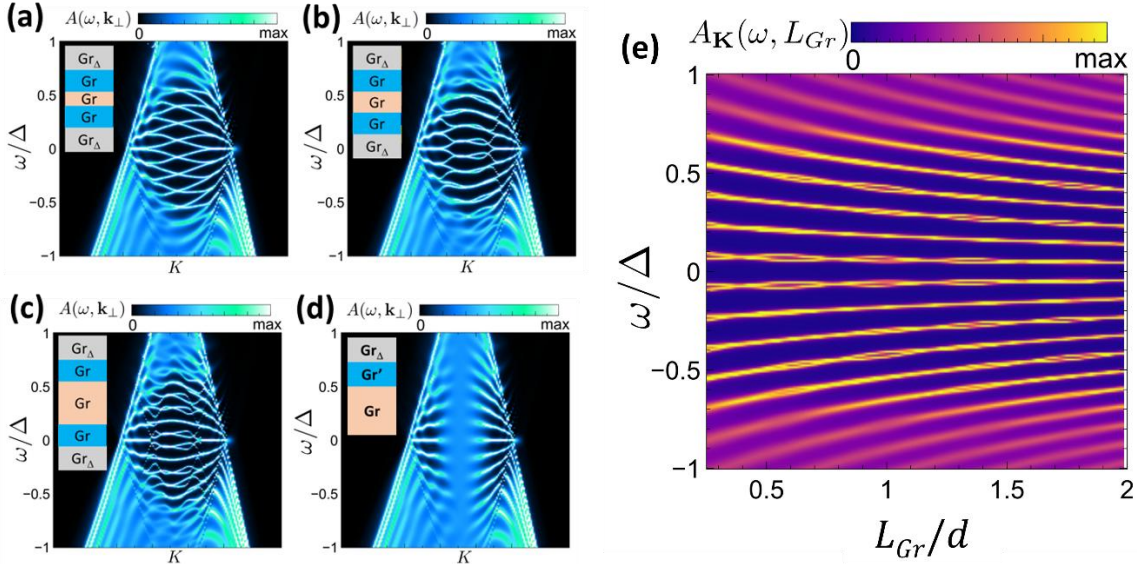

**Figure 3**. (a)-(c): The spectral functions of one $Gr'$ region as the length ratio $L_{Gr}/d$ is increased from ½, 1, and 2 in the $Gr_\Delta - Gr' - Gr - Gr' - Gr_\Delta$ junctions. (d) The spectral function of the $Gr'$ region for a single $Gr_\Delta - Gr' - Gr$ junction. (e) Spectral function at K=0 as a function of $L_{Gr}/d$. The insets in (a)-(d) show the relative geometries of the devices.

**Figure 4** summarizes the results from the experiment and theory. The primary period of the fast Fourier transform (FFT) amplitudes from the experiment (in **Figure 2 (c)**) is plotted as a function of the overall BSCCO-graphene-BSCCO device length ($L$) in **Figure 4**. The vertical and horizontal error bars in the scatter plot correspond to the uncertainty in determining the Dirac point and the variation in the device length across the device. Lack of high bias data on L=0.047 μm device leads to the uncertainty in the Dirac point at low temperature. This uncertainty carries into the FFT analysis and is the reason for its large uncertainty in period shown in **Figure 4**. The period of FFT amplitudes of **Figure 3 (e)** as a function of the normalized junction length ($L_{Gr}/d$) is shown in **Figure 4 (b)**. The secondary peaks in the FFT amplitudes in **Figure 2 (c)** can be attributed to the stray resonances seen at high-momentum in **Figures 3 (b)** and **(c)** and emerge due to increasing secondary scattering within the middle Gr region.

We fit the primary FFT period for experiment and theory as a function of length in **Figure 4** using the power law expression $Al^c$, where $A$ is a pre-factor and $c(<1)$ is the power law factor for a length parameter $l$, which can be the actual length or the normalized length. The fitting parameters for the experimental data are $A \approx 22.15 \pm 1.42$ meV μm and $c = -0.168 \pm 0.025$. In comparison, the fitting parameters for the theoretical primary resonance mode in **Figure 4 (b)** are $A \approx 0.1171 \pm 4.9e-4$ meV μm and $c = -0.253 \pm 6.8e-3$. While an exponential fit works equally well for both data sets (see supplemental Section 6), the discrepancy between the extract parameters is too large.

The relatively small discrepancy between the two fits for the parameter $c$ is attributed to the inevitable disorder present in the experiment but not included in the calculations. Additionally, the lattice size in calculations is much smaller than the actual device size, which limits the number of modes in the $Gr$ that couple the two KARS regions. Nevertheless, the modeling captures the weak power law dependence on length that is observed experimentally.

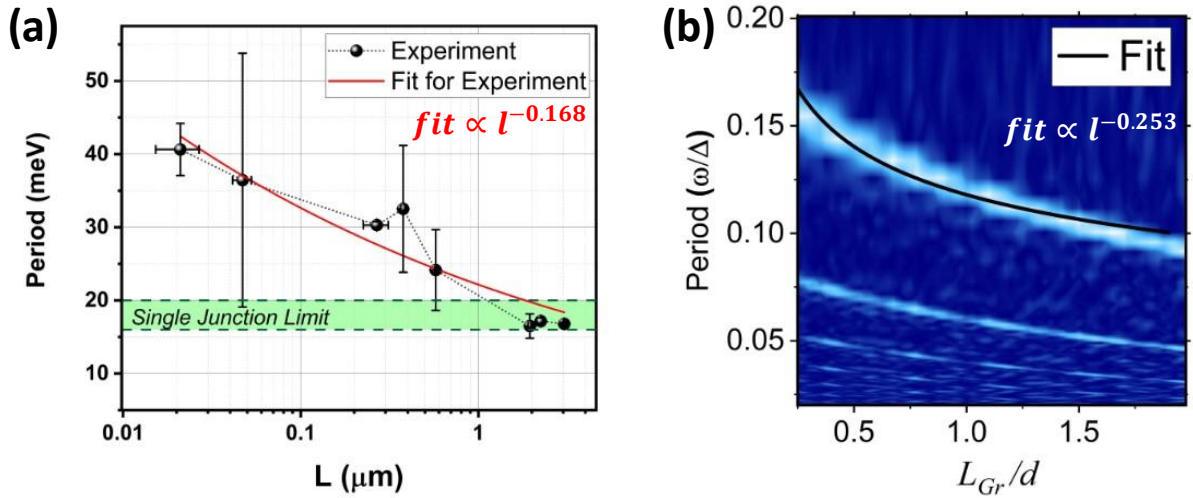

**Figure 4**. (a) The primary period from the experiment plotted as a function of device length (L) is shown in the black scatter plot with error bars. The horizontal error bars indicate the standard error for five data points measured along the width of each device. The error in FFT period (vertical) comes from the uncertainty in the position of the Dirac point. We have assigned the largest error in FFT period for $L = 0.047$ μm because of limited data on $V_{DP}$ at low temperature. The red line is the fit to the experimental data. The range of experimentally measured periods observed in the single BSCCO-Gr'-Gr junctions is highlighted in the green band. The decreasing period of resonances as a function of normalized junction length is shown in (b), generated by taking the Fourier transform of the calculated data in Figure 3 (e). The primary mode follows a power law based on the fit curve (black line) superimposed on the color plot in (b). The fitting parameters for each data set are given in the plots.

In conclusion, we have shown hybridization over long distances between two resonant modes in graphene. Each resonant region is due to KARS that naturally occur at a BSCCO-graphene junction. By forming a BSCCO-graphene-BSCCO double junction with varying separation, we observed hybridization between the two KARS regions. Hybridization occurs over length scales that span over two decades, following a weak power law dependence. Our numerical calculations also support this scaling behavior of the electronic coupling between KARS cavities. Specifically, both experiment and calculations show that the primary period of oscillation scales as $L^c$ power law dependence on length where $c = -0.21 \pm 0.06$. Additionally, both experimental and numerical results show decoupling between the KARS regions as their separation increases, approaching the resonances seen in the single junction. Our results show that novel electronic states with long-range coupling can emerge in S-N-S structures using high-$T_c$ nodal superconductors and graphene. Further understanding and control over such emergent physics in these materials can lead to new devices for quantum sensing [26] and computing [27].

## Associated Content

Supplementary information on bias dependence of resonances and differential conductance can be found in the file 'SGrS_PRX_Supplementary.pdf".

## Author Information

### Corresponding Authors

*Ji Ung Lee – jlee1@albany.edu

*Sharadh Jois – sjois@lps.umd.edu

### Present Address

†Laboratory for Physical Sciences, 8050 Greenmead Dr, College Park, MD 20740

## Data Availability

The data is available in the repository 10.6084/m9.figshare.28282484. Please contact the corresponding authors for additional information.

## Author Contributions

S. Jois and J.U. Lee designed and conceived the experiments. G. Gu and Q. Li grew the BSCCO crystals. S. Jois performed the experimental work and analysis. J. Lado contributed the theoretical figures. S. Jois and J. U. Lee wrote the manuscript.

## Acknowledgement

S. Jois appreciates Gideon Oyibo, Thomas Barret, Mohammad J. Balkan, and Sanjib Banik for assistance during liquid helium transfers. S. Jois and J. U. Lee are thankful to Brian Taylor,

Steve Stewart, and their team for their cooperation and assistance in managing liquid helium orders.

## Funding Sources

This work was supported by funding from the U.S. Naval Research Laboratory Grant No. N00173-19-1-G0008. J. L. L. acknowledges the computational resources provided by the Aalto Science-IT project and the financial support from the Academy of Finland Projects No. 331342 and No. 336243. The work at Brookhaven National Laboratory was supported by the U.S. Department of Energy (DOE) the Office of Basic Energy Sciences, Materials Sciences, and Engineering Division under Contract No. DESC0012704.

# Supplementary Information

*Sharadh Jois[1,†], Jose L. Lado[2], Genda Gu[4], Qiang Li[4,5], *Ji Ung Lee[1,2]

[1] College of Nanoscale Science and Engineering, SUNY Polytechnic Institute, Albany, New York 12203, USA

[2] College of Nanotechnology, Science, and Engineering, University at Albany, SUNY, New York 12203, USA

[3] Department of Applied Physics, Aalto University, 00076 Aalto, Espoo, Finland

[4] Condensed Matter Physics and Materials Science Department, Brookhaven National Laboratory, Upton, New York, 11973, USA

[5] Department of Physics and Astronomy, Stony Brook University, Stony Brook, New York 11794, USA

## 1.   Table of Devices

The **Table S1** below shows the basic properties of each BSCCO-Gr-BSCCO device discussed in the main text in ascending order of device length. The data was reproducible over several devices with similar superconducting transition temperature ($T_c$). The length and width of each device were measured by scanning electron microscopy. Devices on samples S1 through S5 had substrate carrier freeze out for Vbg < 0 V at low temperatures, which we corrected in samples S6 through S9 by increasing the substrate doping. The range for $V_{DP}$ in the table below is based on estimates from data taken at high bias or high temperatures. The length of the Gr' region ($d$), is estimated for long devices with $L > 1$ µm using the relation $d = hv_f/4E_{period}$, where $E_{period}$ is the primary FFT period. $d$ is ill-defined in shorter devices due to hybridization effects.

<u>Note</u>: In failed devices (not included here) the junctions were insulating at room temperature. Creating ohmic interfaces between BSCCO and graphene is achieved by:
- Stacking in an inert nitrogen atmosphere
- Minimizing exposure time (<10 min) of BSCCO's bottom surface after pick-up,
- And maintaining exfoliated graphene sample at room temperature when placing and cracking BSCCO to create the break junctions.

Table S1. Key dimensions of all devices discussed in the manuscript are given below. The estimated length of the Gr' region is $d$ in devices with $L > 1$ µm is also given below. The value of the Dirac point voltage $V_{DP}$ with error range translates to the error bar in FFT period of Figure 4 (a) in the main text.

| Sample#, Dev# | $T_c$ (K) | Device Length (L) (µm) | Width (µm) | $V_{DP}$ (V) | $d$ (nm) | Optical Image |
|---|---|---|---|---|---|---|
| S9, Dev4 | 86 | 0.021 ± 0.006 | 4.6 | -7 ± 1 | NA | 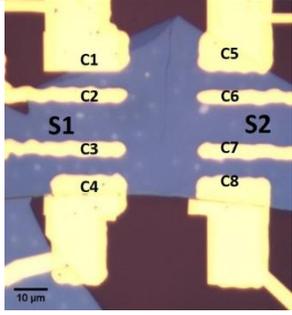 |
| S1, Dev1 | 85 | 0.047 ± 0.006 | 25 | +12.4 ± 2.4 | NA | 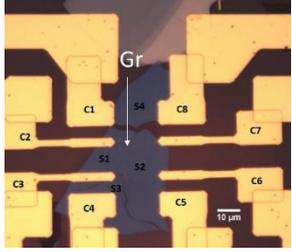 |
| S6, Dev4 | 82 | 0.268 ± 0.043 | 7.8 | +8 ± 0.1 | NA | 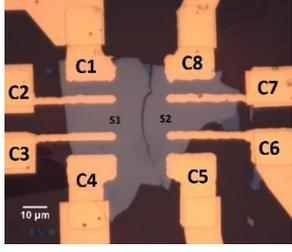 |
| S5, Dev1 | 84 | 0.377 ± 0.011 | 6 | +7.8 ± 1.7 | NA | 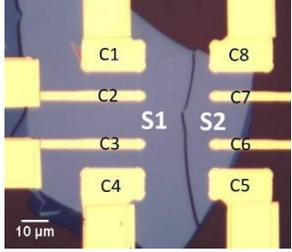 |
| S1, Dev4 | 84 | 0.575 ± 0.006 | 8 | +4.8 ± 1.9 | NA | 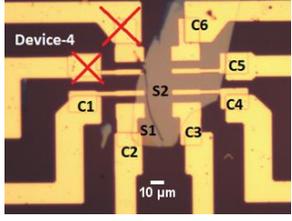 |

| | | | | | |
|---|---|---|---|---|---|
| S1, Dev2 | 79 | 1.946± 0.043 | 6.2 | +3.3 ± 2.5 | 69 | 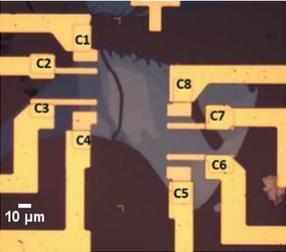 |
| S3, Dev1 | 84 | 2.243± 0.119 | 7.5 | +1.3 ± 0.2 | 66 | 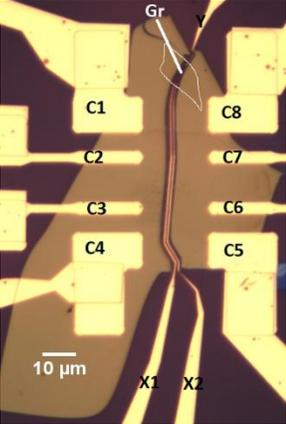 |
| S5, Dev2 | 84 | 3.043± 0.041 | 9 | +6.5 ± 0.8 | 68 | 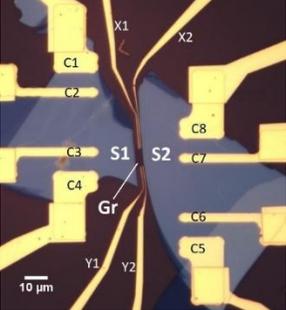 |

## 2. Temperature Dependence of Resonances

In **Figure S1 (a)**, we show the resistance measured as a function of sweeping the back-gate voltage at 5 and 30 K for sample 9, device 4.

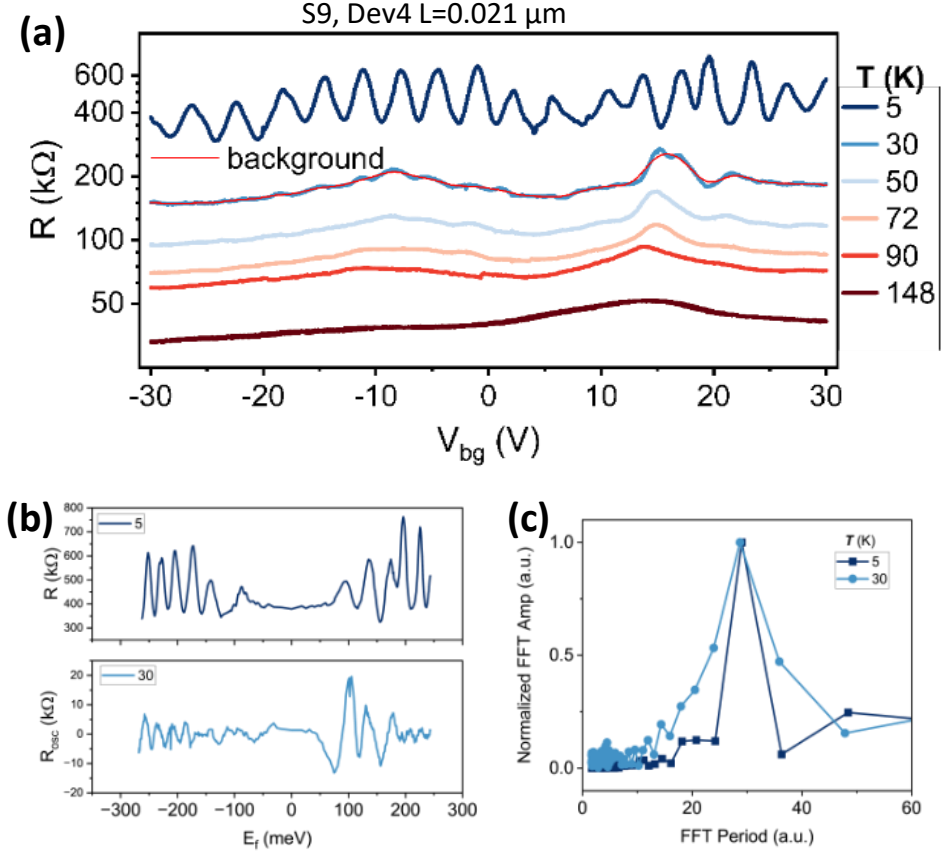

**Figure S1**. Resistance of sample 9, device 4 at different temperatures. In (a) for Sample #9, Device #4, oscillations are seen below 50K. The data at 5 and 30 K with respect to Fermi energy is shown in (b). The oscillatory resistance ($R_{osc}$) is extracted from the raw data at 30 K by subtracting the background (red) and shown in the bottom panel of (b). FFT is taken on the resonances shows a comparable period at both temperatures in (c). At temperatures above 50 K in (a), presence of 2 or 3 Dirac peaks corresponding to different regions of graphene in the devices are identifiable in both devices.

The resonances appear below 50 K. At 30 K and above, we observe two resistance peaks corresponding to the Dirac points of Gr' and Gr regions in the device. The background part of the 30 K data is removed using the FFT or LOESS methods (15-point window) in Origin Lab analysis software, shown in red, to extract the oscillatory component of the data. The raw resistance at 5 K and $R_{osc}$ at 30 K are plotted with respect to Fermi energy in **Figure S1 (b)** and the subsequent FFT analysis are shown in **Figure S1 (c)**. One can clearly observe that the primary period appears unchanged at 5K and 30 K. The oscillations at 50 K are too small in amplitude (<<0.01%) to perform reliable analysis.

## 3. Effects of Back-gate Freeze-Out

In **Figure S2 (a)**, we show the transfer curves for sample S5, Dev1 from room temperature to base liquid nitrogen temperature. The broad peak in resistance evolves into apparent multiple peaks at low temperatures. These peaks are from the different Dirac points pertaining to different regions of graphene in the device. At 90 and 77 K, we see the resistance clips (saturates) abruptly at $V_{bg} \leq 0$ V compared to room temperature data The clipping issue persisted at lower temperatures. This was noticed in samples S1 through S5, and fixed in samples 6-9.

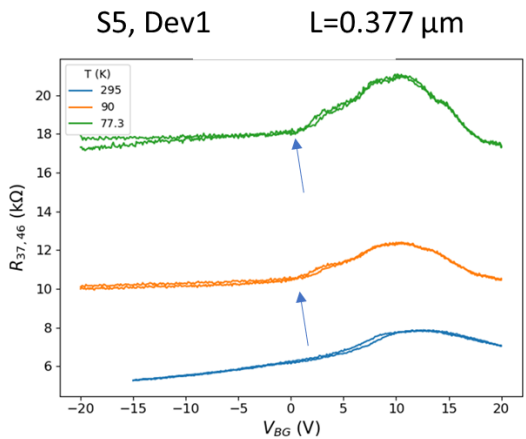

**Figure S2**. (a) Resistance of sample 5, device 1 at different temperatures. Backgate freeze out for Vbg < 0V in Sample S5 Dev-1 at temperatures below 100 K is indicated by the arrow.

## 4. Extracting Oscillatory Component

Below in **Figure S3** we show the raw data for 4 devices from the main text where the oscillatory resistance was extracted for presentation in the main text Figs. 2 (a) and (b). The raw data in these devices can have large background imposed by the modulation of the normal graphene regions. This large background modulation makes it difficult to notice the oscillations and skews the FFT. By examining the FFT of only the oscillatory component of the resistance, we have better confidence in correlating them to the resonances.

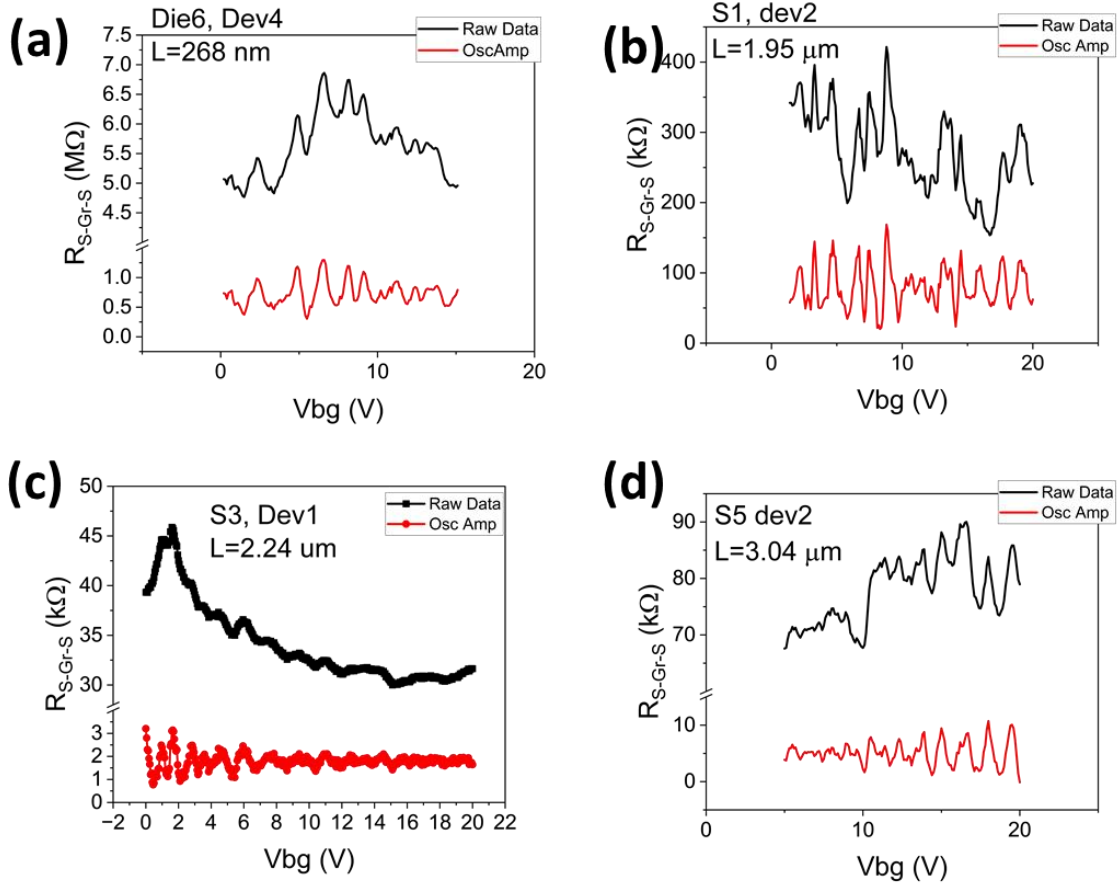

**Figure S3**. Raw data and oscillation amplitude extracted after background subtraction for some devices named in the annotations of figures (a) through (d).

## 5. Bias Dependence

The normalized amplitude of oscillations as a function of source-drain voltage ($V_{ds}$) for the shortest device ($L \approx 21$ nm) is shown in **Figure S4 (a)**. One can see that the oscillation amplitude drops by an order of magnitude when $V_{ds} > 30$ mV. The presence of two superconducting leads in BSCCO-graphene-BSCCO junctions requires twice the voltage of a single junction device to suppress AR. Specifically, the drain bias required to suppress the resonances are $2V_{ds}(S-Gr) = V_{ds}(S-Gr-S) \approx 20\ to\ 30$ meV. In our previous work on single BSCCO-Gr junctions, we found the resonances were suppressed as $V_{ds(S-Gr)} > 10\ to\ 15$ meV, corresponding to the induced superconducting gap in graphene ($\Delta_i$). Thus, the results from these two sets of devices confirm the contribution of Andreev reflection to the resonances.

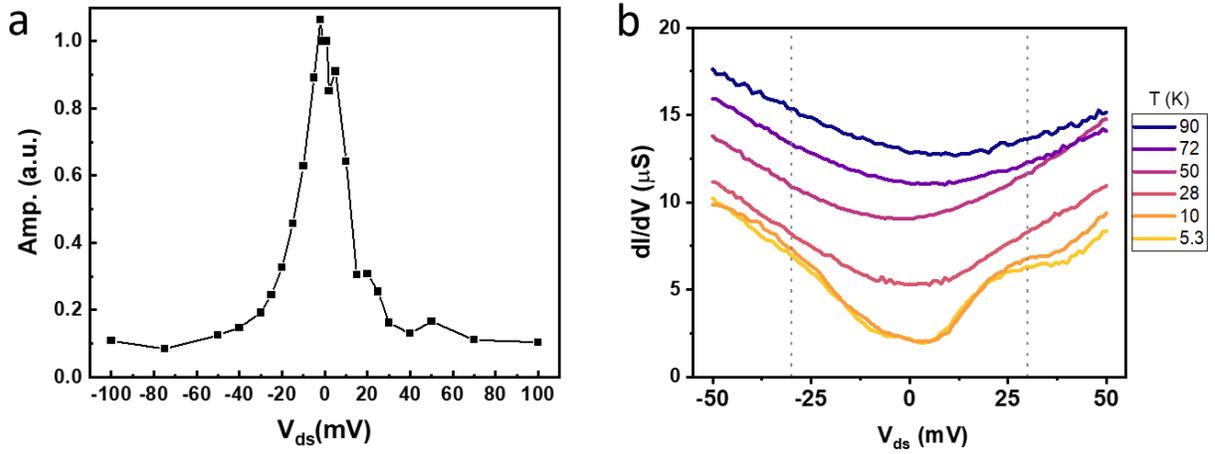

**Figure S4.** The normalized amplitude of oscillations for the shortest device at different DC biases. The oscillation amplitude diminishes at twice the induced gap, $2\Delta_i \sim 30$ meV. (b) The differential conductance (dI/dV) of the S-Gr-S junction as a function of DC bias. A gap-like feature is seen at $2\Delta_i \sim 30$ meV below 10 K.

To further support the induced gap value, we measured the differential conductance across the $L \approx 21$ nm junction of **Figure S4 (a)**. This is shown in **Figure S4 (b)**. The differential conductance was measured at different temperatures and shows the appearance of a gap-like feature below 28 K. Below 10 K, the gap is clearly visible when $V_{ds} \sim 30$ mV and is consistent with the expected value of twice the induced gap $\Delta_i$ we measured for the single junction device. Specifically, the drain bias required to diminish the oscillations in **Figure S4 (a)** corresponds to $V_{ds} \sim 2\Delta_i/e$, consistent with the induced gap $\Delta_i \approx 15$ meV that we measured from our single junction studies and confirms the superconducting origin and the role of AR in these resonances [1]. This result also corroborates with previous reports[2–4]. The correlation between bias dependence of oscillation amplitude and dI/dV of the junction discussed here provides evidence for the role of superconducting contacts in producing the KARS described in the main text.

**Figure S5** shows the oscillation amplitude as a function of applied bias across different BSCCO-Gr-BSCCO (S-Gr-S) devices. The mean fit (line) and standard deviation (band) (in orange) of all the S-Gr-S junctions show that the average bias above which the oscillation amplitude diminishes is $V_{ds}(S - Gr - S) \sim 21$ mV. The same data for single junctions, also shown in **Figure S5**, show that the amplitude diminishes at about half that voltage, $V_{ds}(S - Gr) \sim 9.6$ mV. Lack of high bias data on $L = 0.047$ μm leads to the uncertainty in the Dirac point ($V_{DP}$) at low temperature.

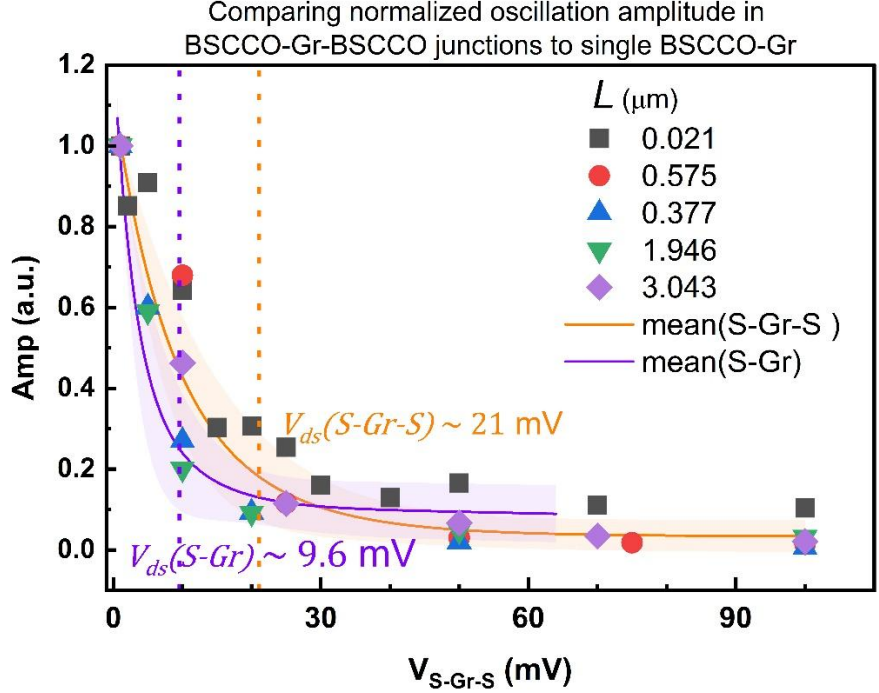

**Figure S5**. Oscillation amplitudes of S-Gr-S and S-Gr devices showing the bias dependence of resonances due to Andreev reflection.

# 6.  Comparing Single S-Gr Junction to S-Gr-S Device

We show in **Figure S6** the comparison between the two individual BSCCO-graphene (S1,2-Gr) junctions and the combined BSCCO-graphene-BSCCO (SGrS) device taken at 4.3 K on Sample #5, Device2 ($L = 3.043$ μm). The raw resistances of the two S-Gr junctions and SGrS device are shown in **Figure S6 (a)**. The background is subtracted from the raw data in **Figure S6 (b)** and plotted as a function of Fermi energy. The FFT result are shown in **Figure S6 (c)**. It shows that the primary peak for the SGrS device (at $14 \pm 2.7$ meV ) lies in between the periods of the individual S1-Gr and S2-Gr junctions ($12.3 \pm 1.5$ and $19.7 \pm 3.1$ meV respectively), confirming that for long devices, the oscillation approach the single junction data. This discussion further helps support the hybridization effects.

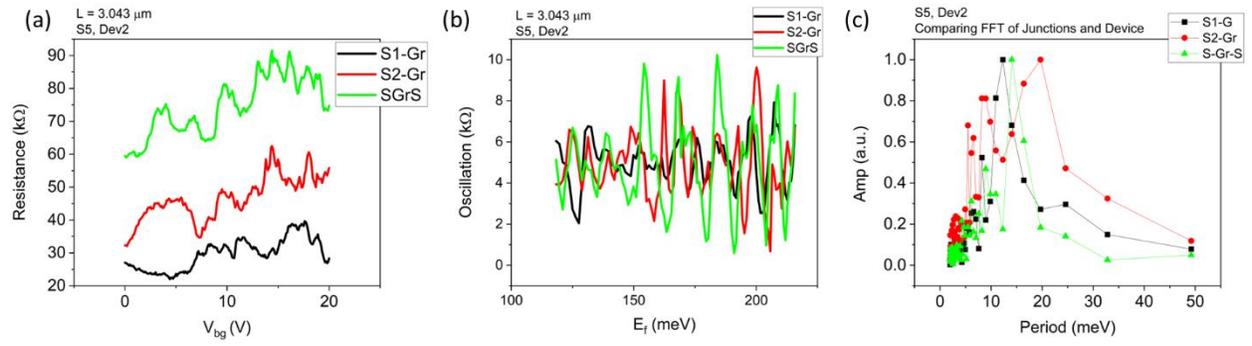

**Figure S6**. (a) Raw resistances of the S1-Gr and S2-Gr junctions are shown alongside the full S-Gr-S device. (b) The oscillatory components associated with resonances in the data in (a) are extracted and shown here as a function of Fermi energy ($E_f$) for direct comparison. (c) FFT amplitudes of the oscillatory resistances in (b) show that the resonance period of S-Gr-S lie between those of S-Gr and S-Gr-S.

# 7. Comparison of Fitting Functions

Due to the significant uncertainty in experiment, we compare fit the primary FFT period for experiment and theory as a function of length in **Figure S7 (a) and (b)** using the power law and exponential functions. The power law fitting function used here is $Al^c$, where $A$ is a pre-factor and $c(<1)$ is the power law factor for a length parameter $l$, which can be the actual length or the normalized length. The exponential fitting function used here, $B\exp^{-\frac{l}{t}} + b_0$, has the pre-factor $B$, denominator in the exponent $t$ and an offset constant $b_0$. Both fits appear to capture the trend seen in the data in **Figure S7 (a) and (b)**. The parameters from both fits are given in the tables of **Figure S7 (c)** and **(d)** for direct comparison. The discrepancy between fit parameters found for experiment and theory can be attributed to the inevitable disorder present in the devices not included in the calculations. Additionally, the lattice size in calculations is much smaller than the actual device size, which limits the number of modes in the $Gr$ that couple the two KARS regions. Upon comparing the differences in fit parameters, we notice a larger difference of the exponential scaling parameter $t$ (factor of 3) between theory and experiment. Whereas the power law scaling parameter between the two cases is more comparable. Although, we cannot decisively say which scaling law governs the coupling between the KARS cavities, we suggest the coupling follows a power law decay based on the agreement in fits.

Owing to the gapless Dirac semi-metal band structure of graphene and appearance of localization phenomena due to electron interference, it seems reasonable that the electronic coupling between the KARS cavities through graphene will obey power laws. Exponential length scaling is commonly attributed to gapped electronic systems.

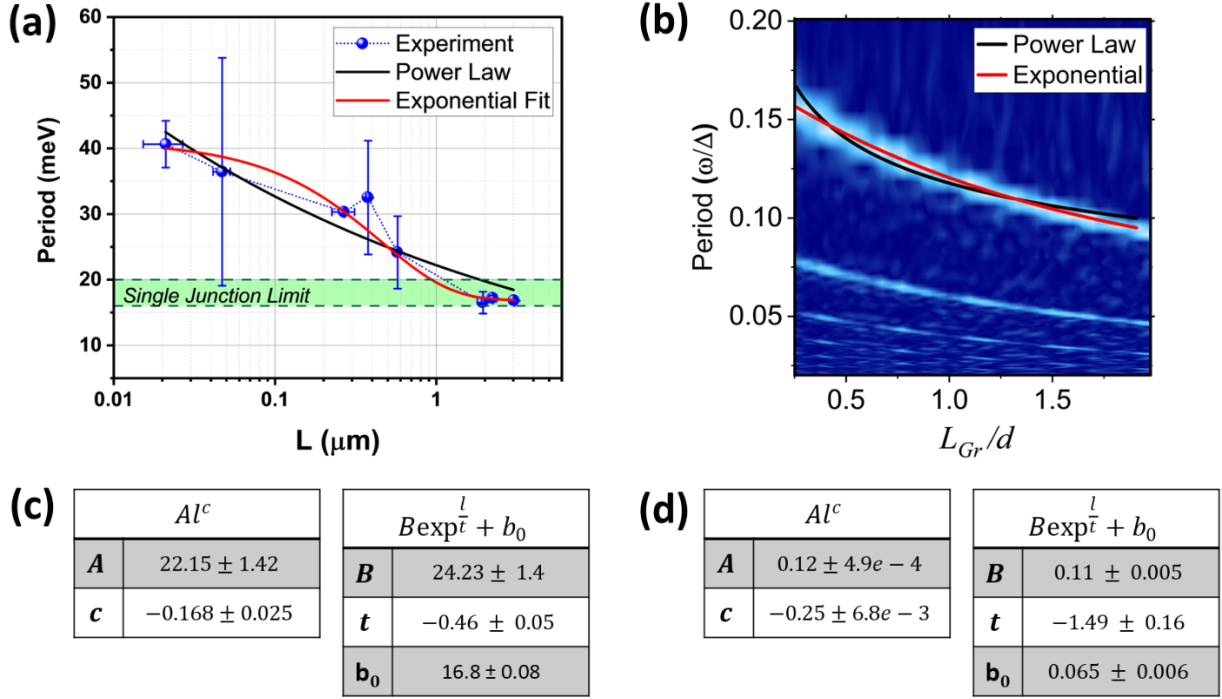

**Figure S7.** Comparison of exponential and power law fits to the experimental (a) and theory (data). The corresponding fit parameters in (c) and (d).

# Supplementary References